\documentclass[prd]{revtex4}
\usepackage{amssymb}
\usepackage{mathptmx}
\usepackage{graphicx}

\newcommand{\shat}{\hat{s}}
\newcommand{\barB}{\overline{B}}
\newcommand{\barK}{\overline{K}}
\newcommand{\kone}{{K_1}}
\newcommand{\barkone}{{\overline{K}_1}}

\newcommand{\pA}{p_{\kone}}
\newcommand{\konel}{K_1(1270)}
\newcommand{\koneh}{K_1(1400)}
\newcommand{\barkonel}{\barK_1(1270)}
\newcommand{\barkoneh}{\barK_1(1400)}
\newcommand{\konea}{K_{1A}}
\newcommand{\koneb}{K_{1B}}
\newcommand{\barkonea}{\barK_{1A}}
\newcommand{\barkoneb}{\barK_{1B}}
\newcommand{\mkone}{m_{\kone}}
\newcommand{\konep}{K_1^+}

\newcommand{\konelm}{K_1^-(1270)}
\newcommand{\konehm}{K_1^-(1400)}

\newcommand{\konelz}{\overline{K}{}^0_1(1270)}
\newcommand{\konehz}{\overline{K}{}^0_1(1400)}
\newcommand{\degree}{^\circ}
\newcommand{\ket}[1]{{#1} \rangle}
\newcommand{\bra}[1]{\langle {#1}}

\newcommand{\sbar}{\bar{s}}

\renewcommand{\l}{\ell}

\newcommand{\lpm}{\l^+\l^-}
\newcommand{\epm}{e^+e^-}
\newcommand{\mupm}{\mu^+\mu^-}
\newcommand{\taupm}{\tau^+\tau^-}
\newcommand{\hats}{\hat{s}}
\newcommand{\hatp}{\hat{p}}
\newcommand{\hatq}{\hat{q}}
\newcommand{\hatm}{\hat{m}}
\newcommand{\hatu}{\hat{u}}
\newcommand{\alphaem}{\alpha_{\rm em}}
\newcommand{\AFB}{A_{\rm FB}}
\newcommand{\barAFB}{\overline{A}_{\rm FB}}
\newcommand{\A}{{\cal A}}
\newcommand{\B}{{\cal B}}
\newcommand{\C}{{\cal C}}
\newcommand{\D}{{\cal D}}
\newcommand{\E}{{\cal E}}
\newcommand{\F}{{\cal F}}
\newcommand{\G}{{\cal G}}
\renewcommand{\H}{{\cal H}}
\renewcommand{\Re}{\mathop{\mbox{Re}}}

\newcommand{\eff}{{\rm eff}}
\newcommand{\GeV}{{\,\mbox{GeV}}}

\newcommand{\Bm}{B^-}
\newcommand{\Bz}{\overline{B}{}^0}
\providecommand{\dfrac}[2]{\frac{\displaystyle
{#1}}{\displaystyle{#2}}}

\begin{document}
\title{$B \to K_1 \ell^+ \ell^-$ Decays in a Family Non-universal $Z^\prime$ Model}
\author{Ying Li$\footnote{liying@ytu.edu.cn}$, Juan Hua}
\affiliation{Department of Physics, Yantai University, Yantai
264-005, China}
\author{Kwei-Chou Yang}
\affiliation{Department of Physics, Chung Yuan Christian University,
Chung-Li, Taiwan 320, Republic of China}
\date{\today}
\begin{abstract}
The implications of the family non-universal $Z^\prime$ model in the
$B\to K_{1}(1270,1400)\ell^{+}\ell^{-}(\ell=e\,,\mu\,,\tau)$  decays
are explored, where the mass eigenstates $K_{1}(1270,1400)$ are the
mixtures of $^{1}{P}_{1}$ and $^{3}{P}_{1}$ states with the mixing
angle $\theta$. In this work, considering the $Z^\prime$ boson and
setting the mixing angle $\theta=(-34\pm13)^{\circ}$, we analyze the
branching ratio, the dilepton invariant mass spectrum, the normalized
forward-backward asymmetry and lepton polarization asymmetries of
each decay mode. We find that all observables of $B\to
K_{1}(1270)\mu^{+}\mu^{-}$  are sensitive to the $Z^{\prime}$
contribution. Moreover, the observables of $B\to
K_{1}(1400)\mu^{+}\mu^{-}$ are relatively strong
$\theta$-dependence; thus, the  $Z^{\prime}$ contribution will be
buried by the uncertainty of the mixing angle $\theta$. Furthermore, the
zero crossing position in the FBA spectrum of $B\to
K_{1}(1270)\mu^{+}\mu^{-}$ at low dilepton mass will move to the
positive direction with $Z^\prime$ contribution. For the tau modes,
the effects of $Z^\prime$ are not remarkable due to the small phase
space. These results could be tested in the running LHC-b experiment
and Super-B factory.
\end{abstract}
\maketitle
\section{Introduction}
The flavor changing neutral currents (FCNC) $b\to s \ell^+
\ell^-(\ell=e,\mu,\tau)$, forbidden in the standard model (SM) at
the tree level, are very sensitive to the flavor structure of the SM
and to the new physics (NP) beyond the SM. The rare decays $B \to
K_1 \ell^+ \ell^-$ involving axial-vector strange mesons, also
induced by $b\to s \ell^+ \ell^-$, have been the subjects of many
theoretical studies in the frame work of the SM
\cite{Hatanaka:2008gu,Li:2009rc,Paracha:2007yx,Bashiry:2009wq} and
some NP models, such as universal extra dimension
\cite{Ahmed:2008ti}, models involving supersymmetry
\cite{Bashiry:2009wh} and the fourth-generation fermions
\cite{Ahmed:2011vr}. Generally, these semileptonic decays provide
us with a wealth of information with a number of physical observables,
such as branching ratio, dilepton invariant mass spectrum, the
forward backward asymmetry, lepton polarization asymmetry and other
distributions of interest, which play important roles in testing SM
and are regarded as probes of  possible NP models.

In the quark model, two lowest nonets of $J^P = 1^+$ axial-vector
mesons are usually expected to be the orbitally excited $q \bar
q\prime$ states. In the context of the spectroscopic notation
$n^{2S+1}L_J$ , where the radial excitation is denoted by the
principal number $n$, there are two types of lowest $p$-wave
meson, namely, $1^{3}P_1$ and $1^{1}P_1$. The two nonets have
distinctive $C$ quantum numbers, $C = +$ or $C =-$, respectively.
Experimentally, the $J^{PC} = 1^{++}$ nonet consists of $a_1(1260)$,
$f_1(1285)$, $f_1(1420)$, and $K_{1A}$, while the $1^{+-}$ nonet
contains $b_1(1235)$, $h_1(1170)$, $h_1(1380)$ and $K_{1B}$. The
physical mass eigenstates $K_1(1270)$ and $K_1(1400)$ are mixtures
of $K_{1A}$ and $K_{1B}$ states owing to the mass difference of the
strange and non-strange light quarks, and the relation could be
written as:
\begin{eqnarray}
\pmatrix{|\barkonel \rangle \cr |\barkoneh \rangle} = M\pmatrix{|
\barkonea \rangle \cr | \barkoneb \rangle},\,\,\,\, \mathrm{with}~~~
M=\pmatrix{ \sin \theta & \phantom{-} \cos \theta \cr \cos \theta &
-\sin \theta}.\label{mixing2}
\end{eqnarray}
In the past few years, many attempts have been made to constrain the
mixing angle $\theta$
\cite{Suzuki:1993yc,Burakovsky:1997ci,Cheng:2003bn,Hatanaka:2008xj}.
In this study, we will use  $\theta= -(34 \pm 13)\degree $for
numerical calculations, which has been extracted from $B\to\konel\gamma$
and $\tau\to\konel\nu_\tau$ by one of us in \cite{Hatanaka:2008xj} ,
and the minus sign is related to the chosen phase of
$|\ket{\barkonea}$ and $|\ket{\barkoneb}$.

To make predictions of these exclusive decays, one requires the
additional knowledge about form factors, i.e., the matrix elements
of the effective Hamiltonian between initial and final states. This
problem, being a part of the nonperturbative sector of QCD, lacks a
precise solution. To the best of our knowledge,  a number of different
approaches had been used to calculate the decay form factors of $B
\to K_1$ decays, such as QCD sum rules \cite{Dag:2010jr}, light cone
sum rules (LCSRs) \cite{Yang:2008xw}, perturbative QCD approach
\cite{Li:2009tx} and light front quark model \cite{Cheng:2009ms}.
Among them, the results obtained by LCSRs which deal with form
factors at small momentum region, are complementary to the lattice
approach and have consistence with perturbative QCD and the heavy
quark limit. On this point, we will use the results of LCSRs
\cite{Yang:2008xw} in this work.

In some new physics models, $Z^\prime$ gauge boson could be
naturally derived in certain string constructions
\cite{Buchalla:1995dp} and $E_6$ models \cite{Nardi:1992nq} by
adding additional $\mathrm{U}(1)^\prime$ gauge symmetry
\cite{Langacker:2000ju}. Among many $Z^\prime$ models, the simplest
one is the family non-universal $Z^\prime$ model. It is of interest
to note that in such a model the non-universal $Z^\prime$ couplings
could lead to FCNCs at tree level as well as introduce new weak
phases, which could explain the CP asymmetries in the current high
energy experiments. The effects of $Z^\prime$ in the $B$ sector have
been investigated in the literature, for example see Ref.
\cite{Barger:2009hn,Cheung:2006tm,Chang:2009wt}. The recent detailed
review is Ref. \cite{Langacker:2008yv}. In Ref. \cite{Chang:2009wt},
Chang {\it et.al} obtained the explicit picture of $Z^\prime$
couplings with the data of $\bar B_s-B_s$ mixing, $B \to K^{(*)}
\ell^+ \ell^-$, $B \to \mu^+\mu^-$, $B \to K\pi$ and inclusive
decays $B \to X_s\ell^+ \ell^-$. So, it should be interesting to
explore the discrepancy of observables between predictions of SM and
those of the  family non-universal $Z^\prime$ model. Motivated by
this, we shall address the effects of the $Z^\prime$ boson in the
rare decays $B \to K_1 \ell^+ \ell^-$.

In experiments, $B \to K_1 \ell^+ \ell^-$ have not yet been measured,
but are expected to be observed at LHC-b \cite{new8} and Super-B
factory \cite{new9}. In particular, it is estimated
that there will be almost $8000$ $B\rightarrow K^{\ast }\mu^+ \mu^-$
events with an integrated luminosity of $2fb^{-1}$ in the LHC-b
experiment \cite{new8,Bettler:2009gt}. Although the branching ratio of $B\rightarrow
K_{1}(1270)\mu^+ \mu^-$ calculated in \cite{Hatanaka:2008gu} is one
order of magnitude smaller than  the experimentally measured value
of $B\rightarrow K^{\ast }\mu^+ \mu^-$ \cite{PDG}, we still expect
the significant number of events for this decay.

The remainder of this paper is organized as follows: in section 2,
we introduce the effective Hamiltonian responsible for the $b\to
s\ell^+\ell^-$ transition in both SM and $Z^\prime$ model. Using the
effective Hamiltonian and $B \to \kone$ form factors, we obtain the
branching ratios as well as various related physical
observables. In section 3, we numerically analyze the considered
observables of $B \to K_1 \ell^+ \ell^-$. This section also includes
a comparison of the results obtained in $Z^\prime$ model with those
predicted by the SM. We will summarize this work in the last
section.
\section{Analytic Formulas }
\subsection{The Effective Hamiltonian for $b \to s$ transition in SM}
By integrating out the heavy degrees of  freedom including top
quark, $W^{\pm}$ and $Z$ bosons above scale $\mu=O(m_b)$, the
effective Hamiltonian responsible for the $b \to s \ell^{+}\ell^{-}$
transitions is given as \cite{Buchalla:1995vs,b to s in theory}:
\begin{eqnarray}
H_{eff}(b\to s \ell^{+}\ell^{-}) &=&
-\frac{G_{F}}{2\sqrt{2}}V_{tb}V_{ts}^{*}
  {\sum\limits_{i=1}^{10}} C_{i}({\mu})O_{i}({\mu}) ,
\label{eff1}
\end{eqnarray}
where we have neglected the terms proportional to $V_{ub}V_{us}^{*}$
on account of $|V_{ub}V_{us}^{*}/V_{tb}V_{ts}^{*}|<0.02$. The
local operators can be found in \cite{Buchalla:1995vs}.
Specifically, the operators $O_9$ and $O_{10}$ are given as
\begin{eqnarray}\label{O910}
O_9=\frac{e^2}{g_s^2}(\bar{s}\gamma_\mu P_Lb)(\bar{\ell}\gamma^\mu
\ell)\,,\quad O_{10}=\frac{e^2}{g_s^2}(\bar{s}\gamma_\mu
P_Lb)(\bar{\ell}\gamma^\mu\gamma_5 \ell)\,.
\end{eqnarray}
In SM, the Wilson coefficients $C_i$ at scale $\mu=m_b$
calculated in the naive dimensional regularization (NDR) scheme
\cite{Buchalla:1995vs} are collected in Table \ref{wilson
coefficient}.
\begin{table}[htbp]
 \begin{center}
 \caption{The SM Wilson coefficients at the scale $\mu=m_b$.}
 \label{wc}
 \vspace{0.3cm}
 \doublerulesep 0.5pt \tabcolsep 0.07in
 \begin{tabular}{lccccccccccc}
 \hline \hline
 $C_1(m_b)$& $C_2(m_b)$& $C_3(m_b)$& $C_4(m_b)$& $C_5(m_b)$& $C_6(m_b)$& $C_7^{\rm eff}(m_b)$& $C_9^{\rm eff}(m_b)-Y(q^2)$& $C_{10}(m_b)$\\\hline
 $-0.274$  & $1.007$   & $-0.004$  & $0.076$   & $0.000$   & $0.001$   & $-0.302$            & $4.094$   & $-4.193$\\
  \hline \hline
 \end{tabular}\label{wilson coefficient}
 \end{center}
 \end{table}
It should be stressed that for $b \to s \ell^{+}\ell^{-}$ processes,
the quark decay amplitude can also receive additional contributions
from the matrix element of four-quark operators, $
\sum\limits_{i=1}^{6}\langle \ell^{+}\ell^{-}s| O_{i}| b \rangle$,
which are usually absorbed into the effective Wilson coefficients.
The effective coefficients $C_{7,9}^{{\rm eff}}$ in Table~\ref{wc}
are defined respectively as~\cite{Buras:1993xp}
\begin{eqnarray}\label{eq:effWC}
&& C_7^{\rm eff} = \frac{4\pi}{\alpha_s}\, C_7 -\frac{1}{3}\, C_3 -
\frac{4}{9}\, C_4 - \frac{20}{3}\, C_5\, -\frac{80}{9}\,C_6\,,
\nonumber\\
&& C_9^{\rm eff} =  \frac{4\pi}{\alpha_s}\,C_9 + Y_{SD}(z,
\shat)+Y_{LD}(z,\shat) ,
\end{eqnarray}
with definitions $z=m_c/m_b, \,\,\, \shat=q^2/m_b^2$.
$Y_{SD}(z,\shat)$ represents the short-distance contributions from
four-quark operators far away from the $c\bar{c}$ resonances
regions, which can be calculated reliably in the perturbative theory. On
the contrary, the long-distance contributions $Y_{LD}(z,\shat)$ from
four-quark operators near the $c\bar{c}$ resonances cannot be
calculated  and are usually parameterized in the form of a
phenomenological Breit-Wigner formula. Currently, the light-cone
distribution amplitudes of the axial-vector mesons actually have not yet been
well studied, since contributions of two axial-vector mesons in the
hadronic dispersion relation cannot be separated in all cases.
Moreover, the width effect of axial-vector meson is so large that
the traditional approach like the sum rules cannot deal with it
effectively. The manifest expressions and discussions for
$Y_{SD}(z,\shat)$ and $Y_{LD}(z,\shat)$, are refereed to
Ref.~\cite{resEff}. Since the contribution of long distance can be
vetoed effectively in the experimental side, we will not discuss it
in the current work. Furthermore, for the $C^{\mathrm{eff}}_7$, we
here also ignore the long-distance contribution of the charm quark
loop, which is suppressed heavily by the Breit-Wigner factor.

\subsection{Family Non-universal $Z^\prime$ Model and Parameter Constraint}
As stated before, in the family non-universal $Z^{\prime}$ model,
there exists the flavor  changing neutral current even at the tree
level due to the non-diagonal chiral coupling matrix. Assuming that
the couplings of right-handed quark flavors with $Z^{\prime}$ boson
are diagonal and ignoring $Z-Z^\prime$ mixing, the $Z^{\prime}$ part
of the effective Hamiltonian for $b\to s l^+ l^-$ can be written as
\cite{Barger:2009hn, Cheung:2006tm, Chang:2009wt}
\begin{equation}\label{Zbsll}
 {H}_{eff}^{Z^{\prime}}(b\to s\ell^+\ell^-)=-\frac{2G_F}{\sqrt{2}}
 V_{tb}V^{\ast}_{ts}\Big[-\frac{B_{sb}^{L}B_{\ell\ell}^{L}}{V_{tb}V^{\ast}_{ts}}
 (\bar{s}b)_{V-A}(\bar{\ell}\ell)_{V-A}-\frac{B_{sb}^{L}B_{\ell\ell}^{R}}{V_{tb}V^{\ast}_{ts}}
 (\bar{s}b)_{V-A}(\bar{\ell}\ell)_{V+A}\Big]+{\rm h.c.}
\end{equation}
To match the effective Hamiltonian in SM, as shown in Eq.(\ref{eff1}),
the above equation is reformulated as
\begin{equation}\label{ZPH}
 {H}_{eff}^{Z^{\prime}}(b\to s\ell^+\ell^-)=-\frac{4G_F}{\sqrt{2}}
 V_{tb}V^{\ast}_{ts}\left[\triangle C_9^{\prime} O_9+\triangle C_{10}^{\prime} O_{10}\right]+{\rm h.c.}\,,
\end{equation}
with
\begin{eqnarray}\label{C910Zp}
 \triangle C_9^{\prime}(m_W)=-\frac{g_s^2}{e^2}\frac{B_{sb}^L
 }{V_{tb}V_{ts}^{\ast}} (B_{\ell\ell}^{L}+B_{\ell\ell}^{R}),\,\,\,\,\,\,\,
 \triangle C_{10}^{\prime}(m_W)= +\frac{g_s^2}{e^2}\frac{B_{sb}^L
 }{V_{tb}V_{ts}^{\ast}} (B_{\ell\ell}^{L}-B_{\ell\ell}^{R}),
\end{eqnarray}
where $B_{sb}^L$ and $B_{ll}^{L,R}$ denote the effective chiral
$Z^{\prime}$ couplings to quarks and leptons. Therefore, the $Z^{\prime}$
contributions can be represented as modifications of the Wilson
coefficients of the corresponding semileptonic operators, i.e.,
$C^{\prime}_{9,10}(m_W)=C_{9,10}^{SM}(m_W)+\triangle
C_{9,10}^{\prime}(m_W)$. The running from $m_{W}$ scale down to
$m_{b}$ is the same as that of SM \cite{Chetyrkin:1996vx,
Altmannshofer:2008dz}, and we had ignored the evolution effect from
$m_{Z^\prime}$ to $m_W$ here. Numerically, with the central values
of the inputs, we get
\begin{eqnarray}
\label{NuVaC9}
 C_9^{\prime}(m_b)=0.0682-28.82\frac{B_{sb}^L
 }{V_{tb}V_{ts}^{\ast}} S_{\ell\ell},\,\,\,\,\,\,\,
 C_{10}^{\prime}(m_b)=-0.0695+28.82\frac{B_{sb}^L
 }{V_{tb}V_{ts}^{\ast}} D_{\ell\ell},
\end{eqnarray}
where $ S_{\ell\ell}=(B_{\ell\ell}^{L}+B_{\ell\ell}^{R})$  and
$D_{\ell\ell}=(B_{\ell\ell}^{L}-B_{\ell\ell}^{R})$.

\subsection{Form Factor}
Following the definitions in Ref. \cite{Yang:2008xw},  the
$\barB(p_B)\to \barkone(\pA,\lambda)$ form factors could be parameterized
as
{\small
\begin{eqnarray}
{\bra{\barkone(\pA,\lambda)}|\sbar \gamma_\mu (1-\gamma_5)
b|\ket{\barB(p_B)}} &=& -i \frac{2}{m_B + \mkone}
\epsilon_{\mu\nu\rho\sigma} \varepsilon_{(\lambda)}^{*\nu} p_B^\rho
\pA^\sigma A^{\kone}(q^2) +2 \mkone \frac{\varepsilon_{(\lambda)}^*
\cdot p_B}{q^2} q_\mu \left[ V_3^{\kone}(q^2) - V_0^{\kone}(q^2)
\right]\nonumber\\
&-& \left[ (m_B + \mkone)\varepsilon_\mu^{(\lambda)*}
V_1^{\kone}(q^2) - (p_B + \pA)_\mu (\varepsilon_{(\lambda)}^* \cdot
p_B) \frac{V_2^{\kone}(q^2)}{m_B + \mkone}
\right],\\\label{formfactor1}
 {\bra{\barkone(\pA,\lambda)}|\sbar
\sigma_{\mu\nu} q^\nu (1+\gamma_5)
b|\ket{\barB(p_B)}}&=&2T_1^{\kone}(q^2) \epsilon_{\mu\nu\rho\sigma}
\varepsilon_{(\lambda)}^{*\nu} p_B^\rho \pA^\sigma-
iT_3^{\kone}(q^2) (\varepsilon_{(\lambda)}^{*} \cdot q) \left[ q_\mu
- \frac{q^2}{m_B^2 - \mkone^2} (\pA + p_B)_\mu \right]
\nonumber\\
&-&i T_2^{\kone}(q^2) \left[
 (m_B^2 - \mkone^2) \varepsilon^{(\lambda)}_{*\mu}
-(\varepsilon_{(\lambda)}^{*}\cdot q)
 (p_B + \pA)_\mu\right], \label{formfactor2}
\end{eqnarray}}
with $q \equiv p_B - \pA$, $\gamma_5 \equiv
i\gamma^0\gamma^1\gamma^2\gamma^3$, and $\epsilon^{0123} = -1$. In the context of
equation of motion, the form factors satisfy the following relations,
\begin{eqnarray}
V_3^\kone(0) &=& V_0^{\kone}(0), \quad T_1^{\kone}(0) =
T_2^{\kone}(0),
\nonumber \\
V_3^\kone(q^2) &=& \frac{m_B + m_{\kone}}{2 m_{\kone}}
V_1^\kone(q^2)-
           \frac{m_B - m_{\kone}}{2 m_{\kone}} V_2^\kone(q^2).
\end{eqnarray}

Because the $\konel$ and $\koneh$ are the mixing states of the
$\konea$ and $\koneb$, the $\barB \to \barK_1$ form factors can be
parameterized as:
\begin{eqnarray}
\pmatrix{
 \bra{\barkonel}|\sbar \gamma_\mu(1-\gamma_5) b |\ket{\barB} \cr
 \bra{\barkoneh}|\sbar \gamma_\mu(1-\gamma_5) b |\ket{\barB}}
&=& M \pmatrix{
 \bra{\barK_{1A}}|\sbar\gamma_\mu(1-\gamma_5) b|\ket{\barB} \cr
 \bra{\barK_{1B}}|\sbar\gamma_\mu(1-\gamma_5) b|\ket{\barB} },
\\
\pmatrix{ \bra{\barkonel}|\sbar \sigma_{\mu\nu}q^\nu(1+\gamma_5) b
|\ket{\barB} \cr \bra{\barkoneh}|\sbar
\sigma_{\mu\nu}q^\nu(1+\gamma_5) b |\ket{\barB} } &=& M \pmatrix{
 \bra{\barK_{1A}}|\sbar\sigma_{\mu\nu}q^\nu(1+\gamma_5) b|\ket{\barB} \cr
 \bra{\barK_{1B}}|\sbar\sigma_{\mu\nu}q^\nu(1+\gamma_5) b|\ket{\barB} }
,
\end{eqnarray}
with the mixing matrix $M$ being given in Eq.~(\ref{mixing2}). Thus
the form factors $A^\kone,V_{0,1,2}^\kone$ and $T_{1,2,3}^\kone$
satisfy following relations:
\begin{eqnarray}
\pmatrix{
 A^{\konel}/(m_B + m_{\konel}) \cr
 A^{\koneh}/(m_B + m_{\koneh})}
&=& M \pmatrix{
 A^{\konea}/(m_B + m_{\konea}) \cr
 A^{\koneb}/(m_B + m_{\koneb})},
\\
\pmatrix{ (m_B+m_{\konel}) V_1^{K_1(1270)} \cr (m_B+m_{\koneh})
V_1^{K_1(1400)}} &=& M \pmatrix{ (m_B+m_{\konea})V_1^{K_{1A}} \cr
(m_B+m_{\koneb})V_1^{K_{1B}}},
\\
\pmatrix{ V_2^{K_1(1270)}/(m_B + m_{\konel}) \cr
V_2^{K_1(1400)}/(m_B + m_{\koneh})} &=& M \pmatrix{
V_2^{K_{1A}}/(m_B + m_{\konea}) \cr V_2^{K_{1B}}/(m_B +
m_{\koneb})},
\\
\pmatrix{ m_{\konel} V_0^{K_1(1270)} \cr m_{\koneh} V_0^{K_1(1400)}}
&=& M \pmatrix{ m_{\konea} V_0^{K_{1A}} \cr m_{\koneb}
V_0^{K_{1B}}},
\\
\pmatrix{ T_1^{K_1(1270)} \cr T_1^{K_1(1400)}}&=& M \pmatrix{
T_1^{K_{1A}} \cr T_1^{K_{1B}} },
\\
\pmatrix{ (m_B^2 - m_{\konel}^2) T_2^{K_1(1270)} \cr (m_B^2 -
m_{\koneh}^2) T_2^{K_1(1400)}}&=& M \pmatrix{ (m_B^2 - m_{\konea}^2)
T_2^{K_{1A}} \cr (m_B^2 - m_{\koneb}^2) T_2^{K_{1B}} },
\\
\pmatrix{ T_3^{K_1(1270)} \cr T_3^{K_1(1400)}} &=& M \pmatrix{
T_3^{K_{1A}} \cr T_3^{K_{1B}} },
\end{eqnarray}
where we have assumed  $p^\mu_{\konel,\koneh} \simeq
p^\mu_{\konea} \simeq p^\mu_{\koneb}$ for simplicity. For the form factors,
we will use results calculated with LCSRs \cite{Yang:2008xw}, which are exhibited in
Table~\ref{tab:FFinLF}. In the whole kinematical region, the dependence of each form factor on momentum
transfer $q^2$  is parameterized in the double-pole form:
\begin{eqnarray} \label{eq:FFpara}
 F(q^2)=\,{F(0)\over 1-a(q^2/m_{B}^2)+b(q^2/m_{B}^2)^2}.
\end{eqnarray}
And the nonperturbative parameters
$a_i$ and $b_i$ can be fitted by the magnitudes of form factors
corresponding to the small momentum transfer calculated in the LCSRs
approach.
\begin{table}[t]
\begin{center}
\caption{Form factors for $B\to K_{1A},K_{1B}$ transitions obtained
in the LCSRs calculation \cite{Yang:2008xw} are fitted to the
3-parameter form in Eq. (\ref{eq:FFpara}).} \label{tab:FFinLF}
\begin{tabular}{clll|clll}
\hline\hline
       $F$
    & $F(0)$
    & $a$
    & $b$
    & $F$
    & $F(0)$
    & $a$
    & $b$
 \\
    \hline
$V_1^{BK_{1A}}$
    & $0.34$
    & $0.64$
    & $0.21$
&$V_1^{BK_{1B}}$
    & $-0.29$
    & $0.73$
    & $0.07$
    \\
$V_2^{BK_{1A}}$
    & $0.41$
    & $1.51$
    & $1.18~~$
&$V_2^{BK_{1B}}$
    & $-0.17$
    & $0.92$
    & $0.86$
    \\
$V_0^{BK_{1A}}$
    & $0.22$
    & $2.40$
    & $1.78~~$
&$V_0^{BK_{1B}}$
    & $-0.45$
    & $1.34$
    & $0.69$
    \\
$A^{BK_{1A}}$
    & $0.45$
    & $1.60$
    & $0.97$
&$A^{BK_{1B}}$
    & $-0.37$
    & $1.72$
    & $0.91$
    \\
$T_1^{BK_{1A}}$
    & $0.31$
    & $2.01$
    & $1.50$
&$T_1^{BK_{1B}}$
    & $-0.25$
    & $1.59$
    & $0.79$
    \\
$T_2^{BK_{1A}}$
    & $0.31$
    & $0.63$
    & $0.39$
&$T_2^{BK_{1B}}$
    & $-0.25$
    & $0.38$
    & $-0.76$
    \\
$T_3^{BK_{1A}}$
    & $0.28$
    & $1.36$
    & $0.72$
&$T_3^{BK_{1B}}$
    & $-0.11$
    & $-1.61$
    & $10.2$
    \\
    \hline\hline
\end{tabular}
\end{center}
\end{table}

\subsection{Formulas of Observables}
With the same convention and notation as \cite{Hatanaka:2008gu},
the dilepton invariant mass spectrum of the lepton pair for the
$\barB\to\barkone\lpm$ decay is given as
\begin{eqnarray}
{\frac{d \Gamma(\barB\to\barkone\lpm)}{d \hats}} &=&\frac{G_F^2
\alphaem^2 m_B^5}{2^{10}\pi^5}
 \left|V_{tb}V_{ts}^*\right|^2 \hatu(\hats)
  \Delta
\end{eqnarray}
and
\begin{eqnarray}
\Delta &=& \frac{\left|\A^\kone\right|^2}{3} \hats \lambda \left(1 +
2\frac{\hatm_\l^2}{\hats}\right) + \left|\E^\kone\right|^2 \hats
\frac{\hatu(\hats)^2}{3} \nonumber\\&+&\frac{1}{4\hatm_\kone^2}
  \Biggl[
     \left|\B^\kone\right|^2 \left(\lambda - \frac{\hatu(\hats)^2}{3}
   + 8\hatm_\kone^2 (\hats + 2\hatm_\l^2 )\right)
+
    \left|\F^\kone\right|^2 \left(\lambda - \frac{\hatu(\hats)^2}{3}
   + 8\hatm_\kone^2 (\hats -4 \hatm_\l^2 )\right)
 \Biggr]
\nonumber\\
&+& \frac{\lambda}{4\hatm_\kone^2} \left[
   \left|\C^\kone\right|^2 \left(\lambda - \frac{\hatu(\hats)^2}{3}  \right)
  +\left|\G^\kone\right|^2 \left(\lambda - \frac{\hatu(\hats)^2}{3}
 + 4\hatm_\l^2 (2 + 2\hatm_\kone^2 - \hats)
  \right)
\right] \nonumber\\&-&
  \frac{1}{2\hatm_\kone^2}
 \Biggl[ \Re \left(\B^\kone \C^{\kone*}\right)
  \left(\lambda - \frac{\hatu(\hats)^2}{3}  \right)
  (1 - \hatm_\kone^2 - \hats)
+\Re \left(\F^\kone \G^{\kone*}\right)
  \left(\left(\lambda - \frac{\hatu(\hats)^2}{3}  \right)
  (1 - \hatm_\kone^2 - \hats)+ 4\hatm_\l^2 \lambda\right)
 \Biggr]
\nonumber\\
&-&2\frac{\hatm_\l^2}{\hatm_\kone^2} \lambda
 \left[
  \Re \left(\F^\kone \H^{\kone*}\right)
  - \Re \left(\G^\kone\H^{\kone*}\right) (1-\hatm_\kone^2)
 \right]
 + \frac{\hatm_\l^2}{\hatm_\kone^2} \hats \lambda \left|\H^\kone\right|^2.
\end{eqnarray}
with $\hatp = p/m_B$, $\hatp_B = p_B/m_B$, $\hatq= q/m_B$,
$\hatm_\kone = m_\kone/m_B$, and $p = p_B + p_\kone$, $q = p_B -
p_\kone = p_+ + p_- $. The auxiliary functions $\A^\kone(\hats),
\cdots, \H^\kone(\hats)$ are defined in Ref.\cite{Hatanaka:2008xj},
and we list them in the Appendix for convenience.  Here, we
also choose $\hats = \hatq^2$ and $\hatu \equiv (\hatp_B -
\hatp_-)^2 - (\hatp_B - \hatp_+)^2$ as the two independent
parameters, which are bounded as $4\hatm_l^2 \le \hats \le
(1-\hatm_{\kone})^2$ and $-\hatu(\hats) \le \hatu \le \hatu(\hats)$,
with $\hatu(\hats) \equiv \sqrt{\lambda
\left(1-{4\hatm_l^2}/{\hats}\right)}$, $\lambda \equiv 1 +
\hatm_{\kone}^2 + \hats^2 - 2\hats - 2\hatm_{\kone}^2 (1+\hats)$.

The differential forward-backward asymmetry of the
$\barB\to\barkone\lpm$ decay is defined as
\begin{eqnarray}
\frac{d \AFB}{d \hats} \equiv
 \int_0^{ \hatu(\hats)} \! d\hatu \frac{d^2 \Gamma}{d\hatu d\hats}
- \int_{-\hatu(\hats)}^0 \! d\hatu \frac{d^2\Gamma}{d\hatu d\hats}.
\end{eqnarray}
Furthermore, the normalized forward-backward asymmetry, which is more
useful in the experimental side, can be written in terms of
quantities as
\begin{eqnarray}
\dfrac{d\barAFB}{d \hats} \equiv \frac{d \AFB}{d \hats} \Bigg/
\frac{d\Gamma}{d \hats}\equiv \hatu (\hats)
   \hats \Big [ \Re\left(\B^\kone \E^{\kone*}\right)
 + \Re\left(\A^\kone \F^{\kone*}\right) \Big ].
\end{eqnarray}
Here, we do not consider the hard spectator corrections since the
light-cone distribution amplitudes of $\kone$ are not precise
enough.

At the end of this section , we pay our attentions on obtaining the lepton polarization asymmetries.
In the center mass frame of dilepton, the three orthogonal unit
vectors could be defined as
\begin{eqnarray}
      \hat{e}_L=\vec{p}_+,\,\,\,\,\,\,
      \hat{e}_N=\frac{\vec{p}_K \times \vec{p}_+ }{|\vec{p}_K \times
      \vec{p}_+|},\,\,\,\,\,\,
      \hat{e}_T= \hat{e}_N \times \hat{e}_L\; ,
\end{eqnarray}
which are related to the spins of leptons by a Lorentz boost. So,
the decay width of the $B \to K_1 \ell^+ \ell^-$ decay for any spin
direction $\hat{n}$ of the lepton, where $\hat{n}$ is a unit vector
in the dilepton center mass frame, can be written as:
\begin{eqnarray}
      \frac{d\Gamma(\hat{n})}{d\hat{s}}=\frac{1}{2}\big (\frac{d\Gamma}{d\hat{s}}\big )_0[1
      +(P_L\hat{e}_L+P_N\hat{e}_N+P_T\hat{e}_T)\cdot\hat{n}].
\end{eqnarray}
In the above equation, the subscript "$0$" denotes the unpolarized
decay width, and $P_L$ and $P_T$  are the longitudinal and transverse
polarization asymmetries in the decay plane, respectively.  $P_N$
is the normal polarization asymmetry in the direction perpendicular
to the decay plane.  Correspondingly, the lepton polarization
asymmetry $P_i ~~(i=L,N,T)$ can be obtained by calculating
\begin{eqnarray}
P_i(\hat{s})=\frac{d\Gamma(\hat{n}=\hat{e}_i)/d\hat{s}-
d\Gamma(\hat{n}=-\hat{e}_i)/d\hat{s}}{d\Gamma(\hat{n}=\hat{e}_i)/d\hat{s}+
d\Gamma(\hat{n}=-\hat{e}_i)/d\hat{s}}\; .
\end{eqnarray}
After a straightforward calculation, we obtain:
 \begin{eqnarray}
 P_L\Delta &=&\sqrt{1-4\frac{\hat{m}^2_l}{\hat{s}}}\left\{\frac{2\hat{s}\lambda}{3}
 \mathrm{Re}(\A\E^{\kone*})+\frac{(\lambda+12\hat{m}^2_{\kone}\hat{s})}{3\hat{m}^2_{\kone}}
 \mathrm{Re}(\B\F^{\kone*})\right.\nonumber\\
 &&\left.-\frac{\lambda(1-\hat{m}^2_{\kone}-\hat{s})}{3\hat{m}^2_{\kone}}\mathrm{Re}(\B\G^{\kone*}+\C\F^{\kone*})
 +\frac{\lambda^2}{3\hat{m}_{\kone}}\mathrm{Re}(\C\G^{\kone*})\right\} ,\\
 P_N\Delta&=&\frac{-\pi\sqrt{\hat{s}}\hat{u}(\hat{s})}{4\hat{m}_{\kone}}\left\{
 \frac{\hat{m}_l}{\hat{m}_{\kone}}\left[\mathrm{Im}(\F\G^{\kone*})(1+3\hat{m}^2_{\kone}-\hat{s})\right.\right. \nonumber\\
 &&\Bigg.\left.+\mathrm{Im}(\F\H^{\kone*})(1-\hat{m}^2_{\kone}-\hat{s})-\mathrm{Im}(\G\H^{\kone*})\lambda  \right]+2\hat{m}_{\kone}\hat{m}_l
 [\mathrm{Im}(\B\E^{\kone*})+\mathrm{Im}(\A\F^{\kone*})]\Bigg\},\\
P_T\Delta&=&\frac{\pi\sqrt{\lambda}\hat{m}_l}{4\sqrt{\hat{s}}}\Bigg\{
4\hat{s}\mathrm{Re}(\A\B^{\kone*})+\frac{(1-\hat{m}^2_{\kone}-\hat{s})}{\hat{m}^2_{\kone}}\left[
-\mathrm{Re}(\B\F^{\kone*})+(1-\hat{m}^2_{\kone})\mathrm{Re}(\B\G^{\kone*})+\hat{s}
\mathrm{Re}(\B\H^{\kone*})\right]\Bigg.\nonumber\\
&&\left.+\frac{\lambda}{\hat{m}^2_{\kone}}[\mathrm{Re}(\C\F^{\kone*})-(1-\hat{m}^2_{\kone})\mathrm{Re}(\C\G^{\kone*})
-\hat{s}\mathrm{Re}(\C\H^{\kone*})]\right\}.
\end{eqnarray}
\section{Numerical Results and Discussion}

\begin{table}[tbp]
\caption{Input parameters}\label{input}
\begin{center}
\begin{tabular}{l}
\hline $m_B=5.279 \GeV$,\quad $\tau_{B^-}=1.638\times 10^{-12}\,{\rm
sec}$,\quad $\tau_{B^0}=1.530\times 10^{-12}\,{\rm sec}$,
\\
$m_{\konel} = 1.272 \GeV$ ,\quad $m_{\koneh} = 1.403 \GeV$,\quad
$m_{\konea} = 1.31 \GeV$ ,\quad $m_{\koneb} = 1.34 \GeV$
\\
 $|V_{tb}^{}V_{ts}^*| = 0.0407$ ,\quad
$m_{b,\rm pole} = 4.8\pm0.2  \GeV $,\quad
  $\alpha_{em} = 1/129$, \quad $\alpha_s(\mu_h) = 0.3$,\\
\hline
\end{tabular}
\end{center}
\end{table}

\begin{table}[t]
\begin{center}
\caption{Predictions for the non-resonant branching fractions
$\mathrm{Br}(B \to \kone\lpm)(10^{-6})$ in the SM and the
non-universal $Z^{\prime}$ model. The first errors  come from the
uncertainty of the $\theta= (-34\pm13)^\circ$ and the second errors
are combination of all uncertainties in the $Z^{\prime}$ model.}
\label{numerical}
\begin{tabular}{ c|ccc|cccccc} \hline \hline
Mode& ~~~~~~~~SM~~~~~~~~& ~~~~~~~~~~~~~~~S1~~~~~~~~~~~~~~~&~~~~~~~~~~~~~~~S2~~~~~~~~~~~~~~~&~~~~Extreme Limit~~~~~ \\
 \hline
$\Bm\to\konelm\epm$     &$24.1^{+0.2}_{-3.6}$  &$33.7^{+0.1}_{-3.5}\pm7.4$ &$28.8^{+0.2}_{-3.0}\pm3.9$ &$49.6^{+0.1}_{-5.3}$ \\%
$\Bm\to\konelm\mupm$    &$19.7^{+0.2}_{-1.8}$  &$29.1^{+0.1}_{-3.5}\pm7.4$ &$24.3^{+0.0}_{-3.2}\pm3.9$ &$44.9^{+0.2}_{-4.2}$ \\%
$\Bm\to\konelm\taupm$   &$0.8^{+0.0}_{-0.2} $  &$0.7^{+0.1}_{-0.1}\pm0.3 $ &$0.8^{+0.0}_{-0.2}\pm0.2 $ &$1.2^{+0.0}_{-0.2} $  \\ %
 \hline
$\Bm\to\konehm\epm$     &$0.9^{-0.4}_{+2.2} $  &$1.2^{-0.4}_{+3.0}\pm0.3$  &$1.0^{-0.3}_{+2.7}\pm0.2$  &$1.6^{-0.4}_{+4.3} $  \\%
$\Bm\to\konehm\mupm$    &$0.5^{-0.0}_{+1.6} $  &$0.8^{-0.0}_{+2.4}\pm0.2$  &$0.7^{-0.1}_{+1.9}\pm0.2$  &$1.3^{-0.1}_{+3.5} $  \\%
$\Bm\to\konehm\taupm$   &$0.01^{-0.00}_{+0.04}$&$0.01^{-0.01}_{+0.04}\pm 0.01$ &$0.01^{-0.01}_{+0.05}\pm 0.01$ &$0.02^{-0.02}_{+0.06}$ \\%
 \hline
$\Bz\to\konelz\epm$     &$22.5^{+0.2}_{-3.4}$  &$31.5^{+0.1}_{-3.3}\pm7.2$ &$26.9^{+0.2}_{-2.8}\pm3.8$ &$46.3^{+0.1}_{-4.6}$ \\%
$\Bz\to\konelz\mupm$    &$18.4^{+0.1}_{-1.7}$  &$27.2^{+0.1}_{-2.5}\pm7.2$ &$22.8^{+0.1}_{-2.2}\pm3.8$ &$41.9^{+0.2}_{-3.9}$ \\%
$\Bz\to\konelz\taupm$   &$0.7^{+0.0}_{-0.1}$   &$0.7^{+0.0}_{-0.1}\pm0.3$  &$0.7^{+0.0}_{-0.1}\pm0.2$  &$1.1^{+0.0}_{-0.1}$  \\%
 \hline
$\Bz\to\konehz\epm$     &$0.8^{-0.3}_{+2.2}$   &$1.1^{-0.4}_{+2.8}\pm0.2$  &$1.0^{-0.4}_{+2.4}\pm0.2$  &$1.5^{-0.3}_{+3.9}$ \\%
$\Bz\to\konehz\mupm$    &$0.5^{-0.0}_{+1.5}$   &$0.8^{-0.1}_{+2.1}\pm0.2$  &$0.6^{-0.0}_{+1.8}\pm0.2$  &$1.2^{-0.1}_{+3.3}$  \\%
$\Bz\to\konehz\taupm$   &$0.01^{-0.01}_{+0.04}$&$0.01^{-0.01}_{+0.04}\pm 0.01$&$0.01^{-0.01}_{+0.04}\pm 0.01$ &$0.02^{-0.02}_{+0.06}$ \\%
  \hline
  \hline
 \end{tabular}
 \end{center}
 \end{table}
\begin{table}[t]
 \begin{center}
 \caption{The inputs parameters for the $Z^{\prime}$ couplings~\cite{Chang:2009wt}. }
 \label{table:parameter}
 \vspace{0.5cm}
 \doublerulesep 0.7pt \tabcolsep 0.1in
 \begin{tabular}{lccccccccc} \hline \hline
  & $|B_{sb}^L|(\times10^{-3})$ & $\phi_{s}^L[^{\circ}]$ & $B^{L}_{\mu\mu}(\times10^{-2})$ & $B^{R}_{\mu\mu}(\times10^{-2})$\\\hline
 S1 & $1.09\pm0.22$ & $-72\pm7$   & $-4.75\pm2.44$    & $1.97\pm2.24$   \\
 S2 & $2.20\pm0.15$ & $-82\pm4$   & $-1.83\pm0.82$    & $0.68\pm0.85$   \\
  \hline \hline
 \end{tabular}
 \end{center}
 \end{table}
\begin{figure}
\begin{center}
\includegraphics[scale=1]{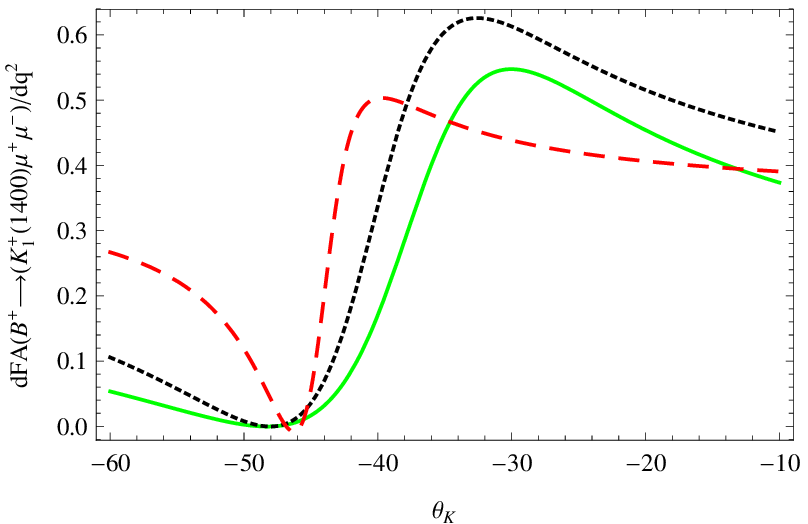}~~~~~~~
\includegraphics[scale=1]{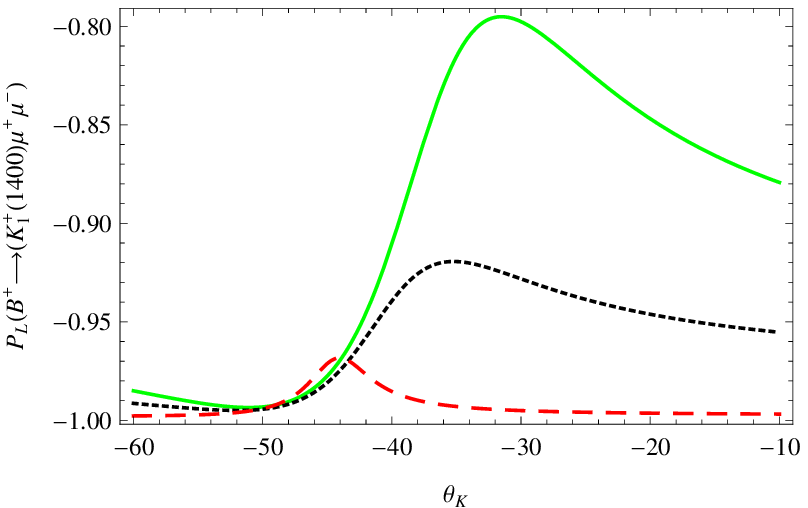}
\caption{Normalized differential forward-backward asymmetries (left
panel) and longitudinal lepton polarization asymmetry(right panel)
of $B\to\koneh \mupm$, as a function of $\theta$(in units of
degree). The solid, dotted and dashed curves correspond to $s=
5\GeV^2$, $7\GeV^2$and $12\GeV^2$, respectively.}\label{Fig:1}
\end{center}
\end{figure}
\begin{figure}
\begin{center}
\includegraphics[scale=1]{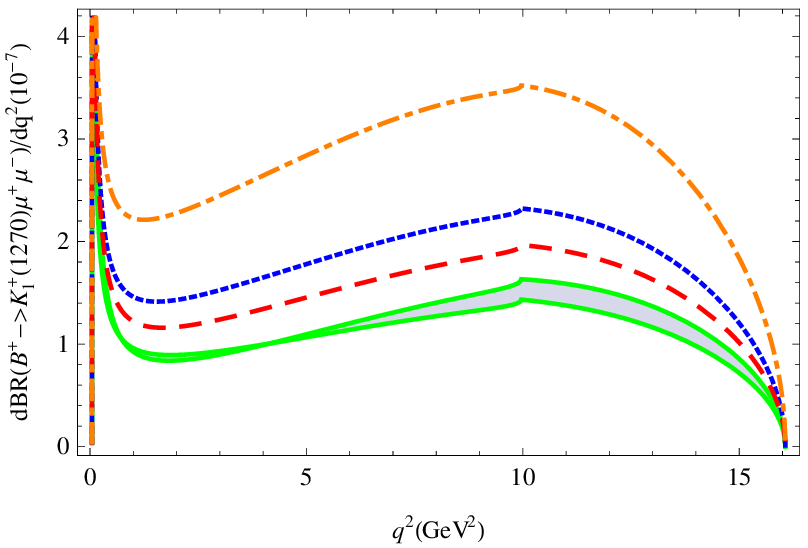}~~~~~~~
\includegraphics[scale=1]{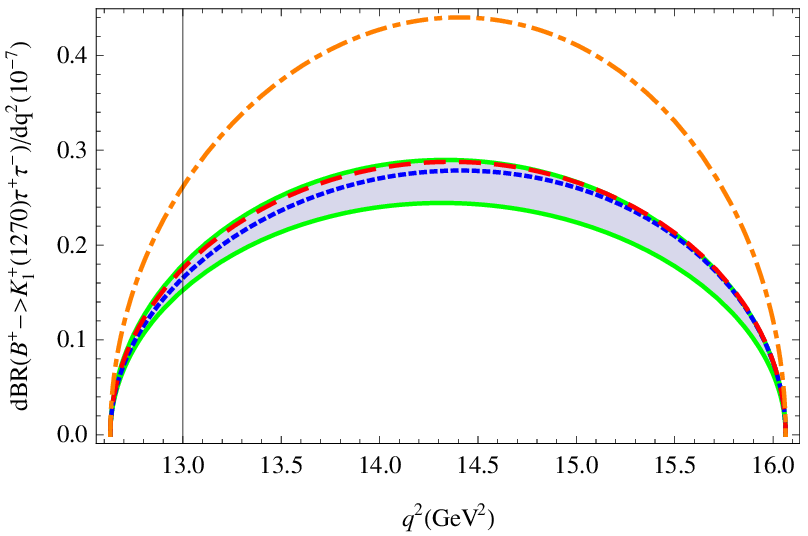}
\includegraphics[scale=1]{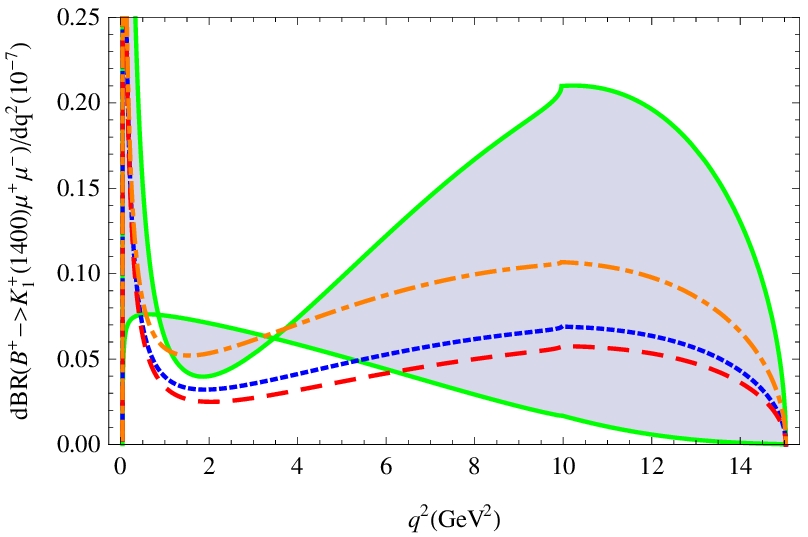}~~~~~~~
\includegraphics[scale=1]{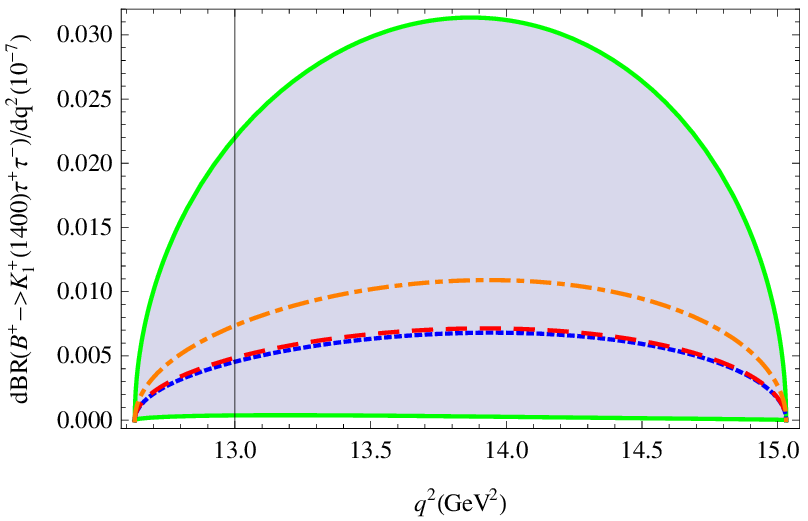}
\caption{The differential decay rates $d Br(B^+\to\konep \lpm)/dq^2$
as functions of $q^2$(in units of $\mathrm{GeV}^2$). The central
values of inputs are used. The solid (green), and dotted (blue),
dashed (red) and dot-dashed (orange) lines represent results from
the standard model, S1, S2 and ELV parameters,
respectively.}\label{Fig:2}
\end{center}
\end{figure}
\begin{figure}
\begin{center}
\includegraphics[scale=1]{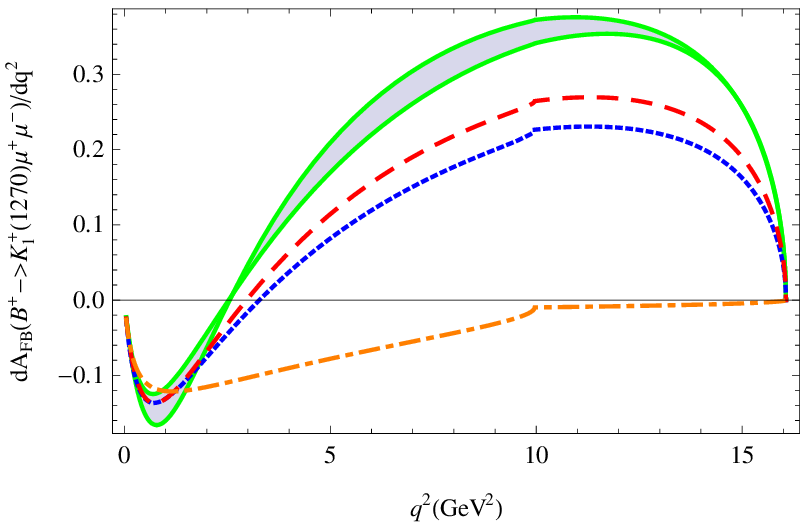}~~~~~~~
\includegraphics[scale=1]{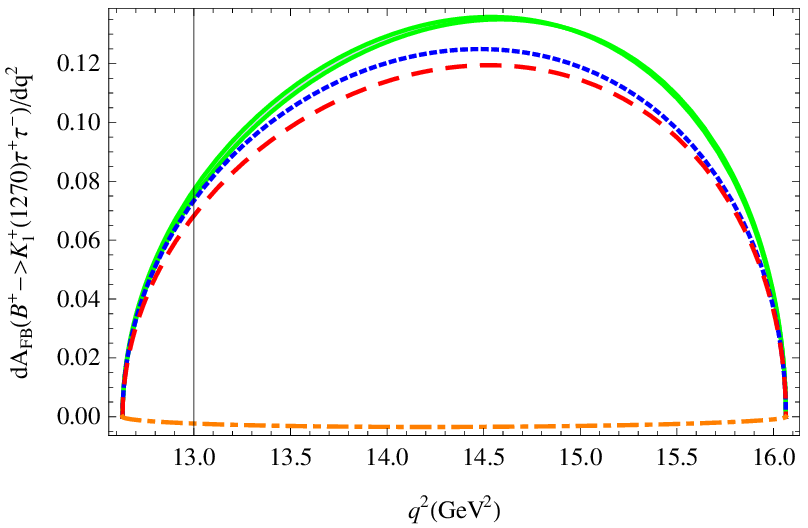}
\includegraphics[scale=1]{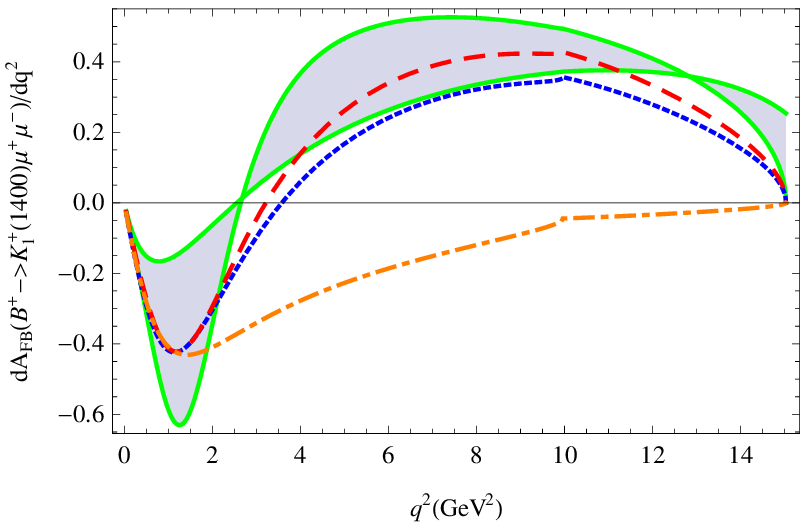}~~~~~~~
\includegraphics[scale=1]{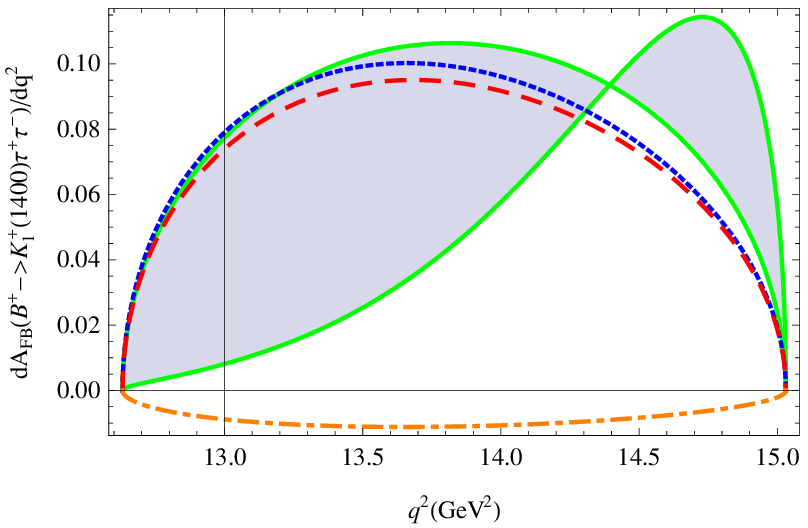}
\caption{The normalized differential forward-backward asymmetries
for the $B^+\to\konep \lpm$ decays as functions of $q^2$(in units of
$\mathrm{GeV}^2$).}\label{Fig:3}
\end{center}
\end{figure}
\begin{figure}
\begin{center}
\includegraphics[scale=1]{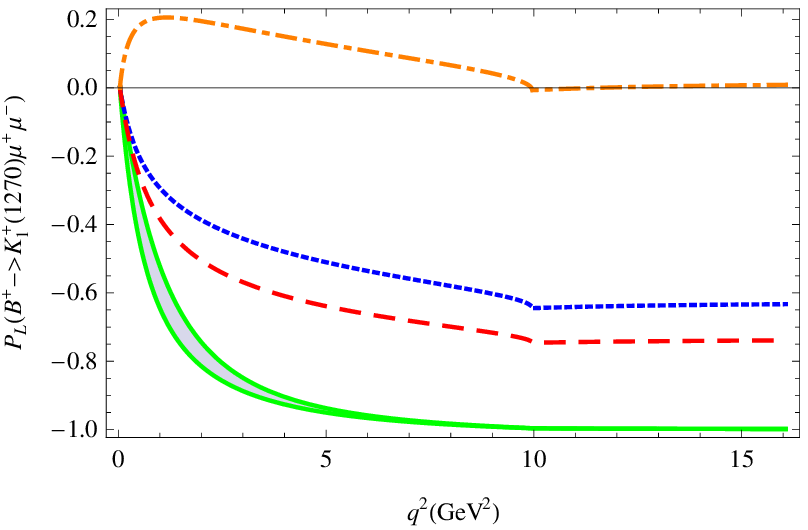}~~~~~~~
\includegraphics[scale=1]{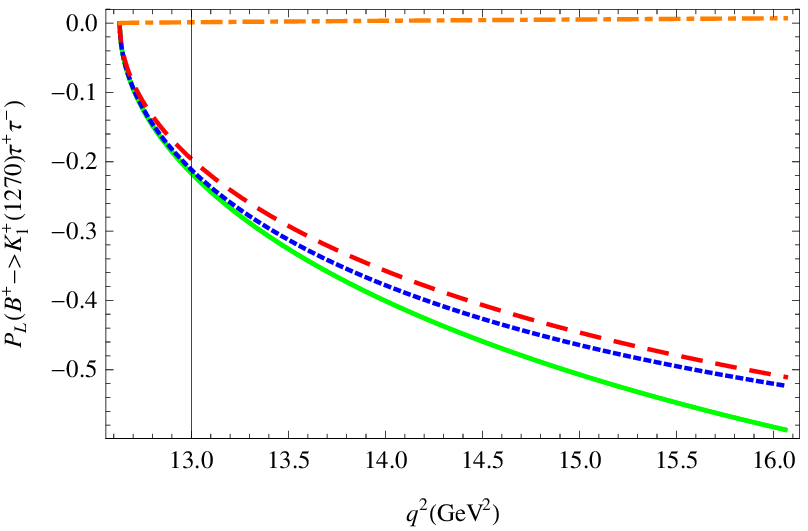}
\includegraphics[scale=1]{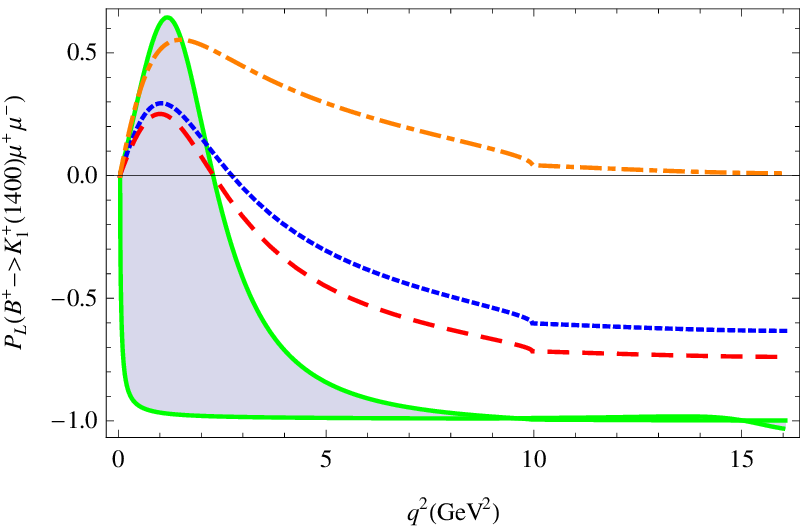}~~~~~~~
\includegraphics[scale=1]{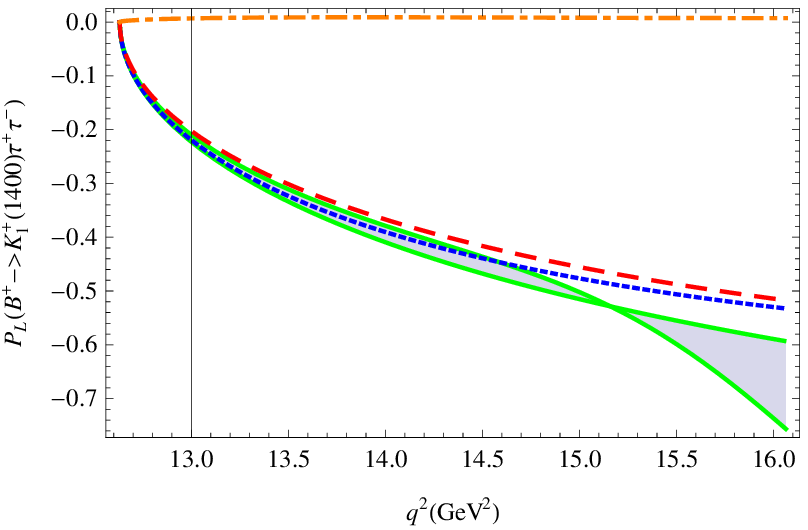}
\caption{The longitudinal lepton polarization asymmetries for the
$B^+\to\konep \lpm$ decays as functions of $q^2$(in units of
$\mathrm{GeV}^2$).}\label{Fig:4}
\end{center}
\end{figure}
\begin{figure}
\begin{center}
\includegraphics[scale=1]{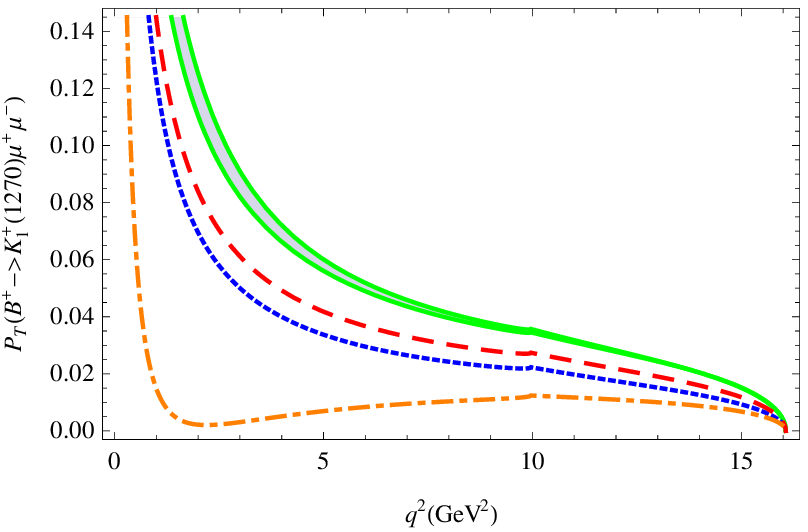}~~~~~~~
\includegraphics[scale=1]{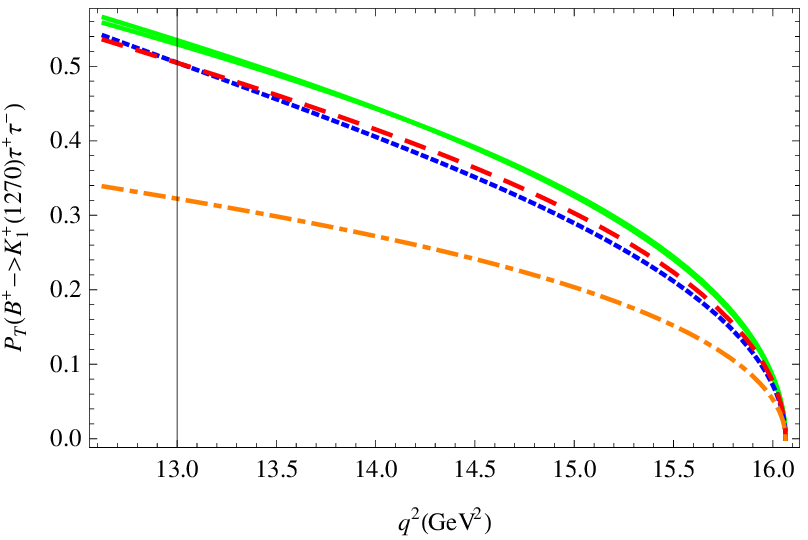}
\includegraphics[scale=1]{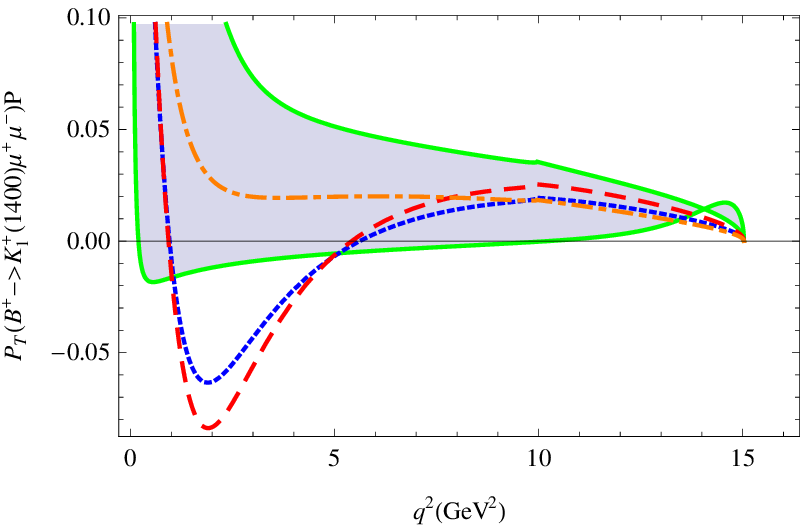}~~~~~~~
\includegraphics[scale=1]{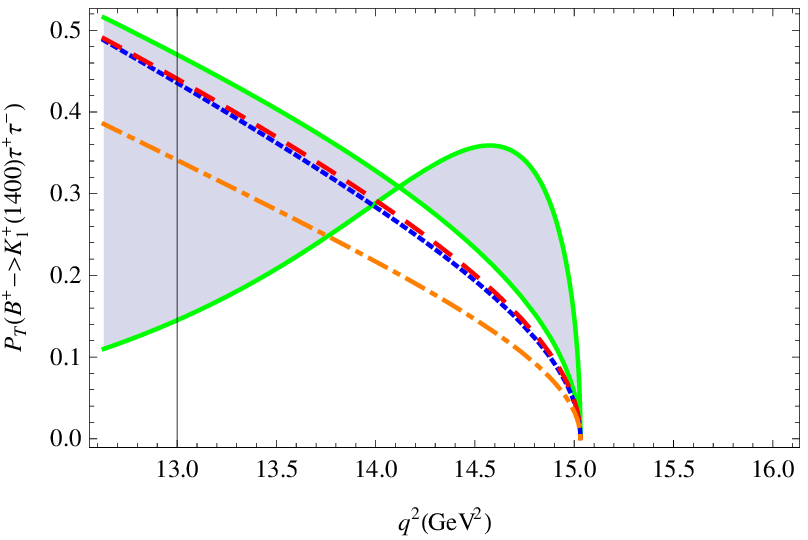}
\caption{The transverse lepton polarization asymmetries for the
$B^+\to\konep \lpm$ decays as functions of $q^2$(in units of
$\mathrm{GeV}^2$).}\label{Fig:5}
\end{center}
\end{figure}
\begin{figure}
\begin{center}
\includegraphics[scale=1]{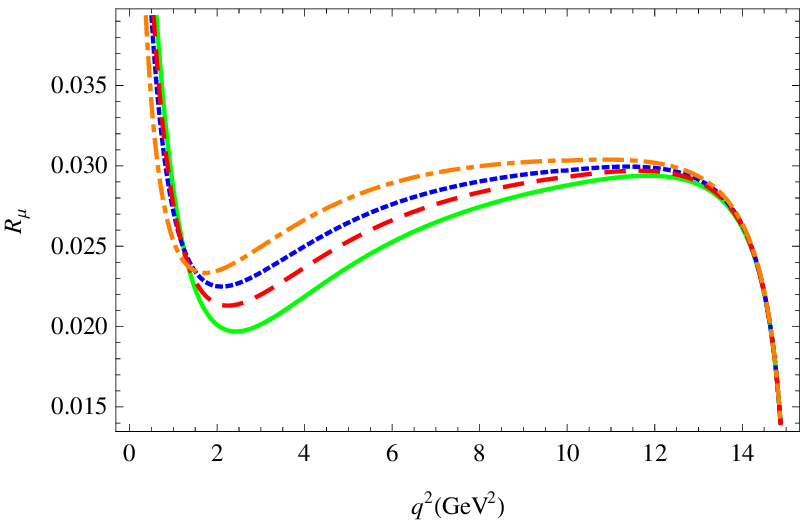}~~~~~~~
\includegraphics[scale=1]{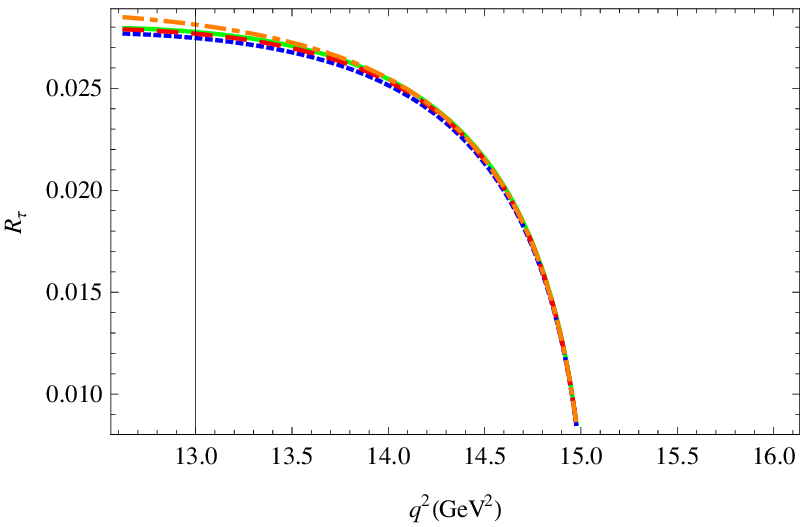}
\caption{The ratio of the decay distributions, $R_\mu$ (left panel)
and $R_\tau$ (right panel), as a function of the dilepton invariant
mass $q^2$(in units of $\mathrm{GeV}^2$).  The legends are the same
as in Fig.~\ref{Fig:2}}\label{Fig:6}
\end{center}
\end{figure}
In this section, we shall calculate aforementioned
observables like the branching ratios (BR), the normalized forward-backward
asymmetries (FBA) and lepton polarization asymmetries, as well as
their sensitivities to the new physics due to $Z^\prime$ boson. The input parameters
used in the numerical calculations are listed in Table.\ref{input}. In discussing
the $\konel$ and $\koneh$, we have to draw much attention on the
mixing angle $\theta$ defined in Eq.(\ref{mixing2}), although many
attempts have been done to constrain it. The magnitude of $\theta$
was estimated to be $|\theta| \approx 34\degree\vee 57\degree$ in
Ref. \cite{Suzuki:1993yc}, $35\degree \leq |\theta| \leq 55\degree$
in Ref.~\cite{Burakovsky:1997ci}, and $|\theta|= 37\degree \vee
58\degree$ in Ref.~\cite{Cheng:2003bn}. Nevertheless, the sign of
the $\theta$ was yet unknown in above studies. From the studies of
$B\to\konel\gamma$ and $\tau\to\konel\nu_\tau$, one of us recently
obtained \cite{Hatanaka:2008xj}
\begin{eqnarray}
\theta= -(34 \pm 13)\degree, \label{thetaKvalue}
\end{eqnarray}
where the minus sign of $\theta$ is related to the chosen phase of
$|\ket{\barkonea}$ and $|\ket{\barkoneb}$, and we will use this
range in the following discussion.

In Ref.~\cite{Hatanaka:2008gu}, the authors found that in the low
$s$ region, where $s\approx 2\GeV^2$, the differential decay rate
for $B\to K_1(1400) \mu^+ \mu^-$ with $\theta=-57^\circ$ is enhanced
by about 80\% compared with that with $\theta=-34^\circ$, whereas
the rate for $B\to K_1(1270) \mu^+ \mu^-$ is not so sensitive to
variation of $\theta$. After calculation, we emphasize that all
observables of $B\to K_1(1400) \mu^+ \mu^-$ are sensitive to the
mixing angle. In Fig.\ref{Fig:1}, for instance, we plot the
relations of the normalized forward-backward asymmetry and longitudinal lepton
polarization asymmetry of $B\to\koneh \mupm$ with $\theta$ varying
from $-10^\circ$ to $-60^\circ$, when $s= 5\GeV^2$ (solid line),
$7\GeV^2$ (dotted line) and $12\GeV^2$ (dashed line). With these
figures and data, one can constrain the angle  in future, as well as
cross check the bands from other theories and experiments.
Additionally, because of small masses of electron and muon, the
invariant mass spectra and branching ratios are almost the same for
electron and muon modes. Meanwhile, it is very difficult to measure
the electron polarization, so we only consider $B \to \kone
\mu^+\mu^-, \kone \tau^+\tau^-$ except for numerical results in the
following discussions.

In Table~\ref{numerical}, we again summarize the predictions for
branching fractions corresponding to $\theta=-(34\pm13)\degree$
without considering the uncertainties taken by the form factors, which
have been discussed in detail in Ref. \cite{Hatanaka:2008gu}. The
negligible disparities between our results and those of
Ref.\cite{Hatanaka:2008gu} are from the difference of Wilson
coefficients. From the table, we note that the branching ratios of  $B\to \konel
\ell^+\ell^-$ are not sensitive to the mixing angle $\theta$, while
those of $B\to \koneh \ell^+\ell^-$ are sensitive to it seriously. We also
find that the branching ratios of $B\to \konel \ell^+\ell^-$ are
much larger than those of $B\to \koneh \ell^+\ell^-$. For $B\to \kone
\tau^+\tau^-$, the branching ratios are very small due to small
phase spaces.

Now, we turn to a discussion of the new physics contribution. Within a
family non-universal $Z^{\prime}$ model, the $Z^{\prime}$
contribution to $B\to \kone \ell^+\ell^-$ decay involves four new
parameters $|B_{sb}^L|$, $\phi_{s}^L$,
$B^{L}_{\ell\ell}$ and $B^{R}_{\ell\ell}$. The tasks of constraining
the above parameters from the well measured channels have been done
by many groups in the past few years. Combining the constraints from
$\bar{B}_s-B_s$ mixing, $B\to\pi K^{(\ast)}$ and $\rho K$ decays,
$|B_{sb}^L|$ and $\phi_{s}^L$ have been strictly constrained by
Chang {\it et. al.}~\cite{Chang:2009wt}. They also performed the
constraints of $B_{ll}^{L,R}$ from $B\to X_s\mupm$, $K\mupm$ and
$K^{\ast}\mupm$, as well as $B_s\to\mupm$ decay. Recently, there
have been more data from Tevatron and LHC on decay processes
mentioned above. Many of them might afford stronger upper bounds
than before, but the new parameters have not been fitted and we will
leave it as our future work. In the current work, we will adopt the
parameters fitted in Ref~\cite{Chang:2009wt} so as to probe
contribution of new physics with the largest possibility. For
convenience, we recollect their numerical results in the
Table~\ref{table:parameter}, where S1 and S2 correspond to UTfit
collaboration's two fitting results for $\bar{B}_s-B_s$
mixing~\cite{UTfit}. Meanwhile, in order to show the maximal
strength of $Z^\prime$, with permitted range in S1, we choose the
extreme values
 \begin{eqnarray}
 |B_{sb}^L|=1.31\times10^{-3}, \phi_{s}^L=-79^{\circ}, S_{ll}=
-6.7\times10^{-2}, D_{ll}=-9.3\times10^{-2},
 \end{eqnarray}
and name them as extreme limit values (ELV) expediently.

Considering the $Z^\prime$ contribution with two sets of parameters,
we calculate the non-resonant branching ratios of concerned decay
modes and tabulate them in the third and fourth columns of the
Table.\ref{numerical}, where the first errors come from the mixing
angle $\theta$ and the second errors are from all uncertainties of
$Z^{\prime}$ model by adding all the theoretical errors in
quadrature.  With the ELV parameters, the extreme results are listed
in the last column of Table.\ref{numerical}, and the errors are only
from mixing angle.

In Fig.\ref{Fig:2}-\ref{Fig:5}, we plot  the differential
branching ratios, forward-backward asymmetries and polarization
asymmetries of the leptons of $B^+\to \konel^+ \ell^+\ell^-$ and
$B^+\to \koneh^+ \ell^+\ell^-$, respectively. In all figures, the
bands with solid (green) lines are results from the standard model
with  $\theta= -(34 \pm 13)\degree$, and dotted (blue), dashed (red)
and dot-dashed (orange) lines represent the results with the S1, S2
and ELV parameters by fixing $\theta= -34\degree$, respectively.
Some discussions of the above results are in order.

\begin{itemize}
\item From the Table. \ref{numerical}, we find that the effect of S1 is more significant
than that of S2. For the central values, compared with predictions
of the SM, the branching ratio of $B^- \to \konel^- \mu^+\mu^-$ can
be enhanced about by $48\%$ in S1, and by $23\%$ in S2. If we choose the
extreme limit of S1, the branching ratio can be enhanced one time at
most by new physics contribution of $Z^\prime$. As concerns $B\to \koneh
\ell^+\ell^-$, their branching ratios are more
sensitive to the mixing angle than to a new physics contribution, and then it
is very hard to differentiate the $Z^\prime$ effects.

\item For the dilepton invariant mass spectrum of
$B^- \to \konel^- \mu^+\mu^-$, the effects of the $Z^\prime$ boson are
quite distinctive from that of the SM, as shown in Fig.\ref{Fig:2}. The reason for the
enlargement is the relative change of the absolute values of
$C_{9}^{\mathrm{eff}}$ and $C_{10}$, though the latter is $q^2$
independent. For $B^- \to \koneh^- \mu^+\mu^-$, $Z^\prime$
boson could change the shape effectively, however this contribution
would be clouded by the uncertainties from the mixing angle. For the
tauon modes, with large tauon mass and small phase space, it is very
difficult to disentangle the new physics contribution from the
predictions of SM, unless choosing the extreme limit case.

\item We plot the normalized forward-backward asymmetries in
Fig.\ref{Fig:3}. For $B\to \kone \mupm$, there  exist zero
crossing positions in SM, S1 and S2. We would like to emphasize that
the hadronic uncertainties and mixing angle almost have no influence
on zero crossing positions, as shown in figures. Specifically, for $B\to
\konel\mu^+\mu^-$, the zero crossing $s_0$ positions are $2.3~\mathrm{GeV}$,
$3.3~\mathrm{GeV}$ and $2.9~\mathrm{GeV}$ in SM, S1 and S2.
Accordingly, for $B\to \koneh\mu^+\mu^-$, $s_0= 2.8~\mathrm{GeV},3.5~
\mathrm{GeV}, 3.2~\mathrm{GeV}$. It is obvious that $s_0$ moves to
the positive direction with the $Z^\prime$ boson effects. And in the limit
values, the zero crossing positions disappear. Thus, the measurement of zero
position is very important for searching for new physics
contribution in the experiments. For $B\to \koneh\tau^+\tau^-$, with the central value
of S1 and S2, the forward-backward asymmetries are almost the same as the
predictions from SM. However, these asymmetries become almost zero in
both low and large momentum regions in the ELV case.

\item Just like the $BR$ and $A_{FB}$, the polarization asymmetries of leptons are also
good observables for probing the new physics signals. In order to
show the effects due to the $Z^\prime$, we figure out the
longitudinal $P_L$ and transverse $P_T$ polarization asymmetries as
functions of $q^2$ in Fig.\ref{Fig:4} and Fig.\ref{Fig:5},
respectively.  The $P_N$ parts are too tiny to be
measured experimentally even in the designed Super-B factory,
so we will not discuss them in this work.
In the case of $B \to \konel \mu^+\mu^-$, the longitudinal
(transverse) polarization asymmetry of lepton is enhanced
(decreased) remarkably by new physics effects. In SM, the longitudinal
polarization asymmetry for muon is around $-1$ in the large momentum
region, while it changes to $-0.6 (-0.7)$ in S1 (S2). In the $Z^\prime $ model,
a large value of differential decay rate  will suppress the absolute value of longitudinal
polarization asymmetry in the large $q^2$ part. With the extreme values,
$P_L$ flips the sign in the low $q^2$ region and approaches to zero
in the large $q^2$ region. If there exist large $Z^\prime-b-s$
and $Z^\prime-l-l$ couplings, we can check them by measuring
the above observables. Similar effects can be found in tau modes, but the deviations are too
small to be measured experimentally.

\item From the Table.\ref{numerical}, we obtain
$Br(B \to \konel \lpm) \gg  Br(B\to\koneh\lpm)$. It should be
helpful to define the ratio $R_{\l}$, as mentioned in
Ref.\cite{Hatanaka:2008gu},
\begin{eqnarray}
R_ \l \equiv
\frac{d\Gamma(\Bm\to\konehm\lpm)/ds}{d\Gamma(\Bm\to\konelm\lpm)/ds}.
\end{eqnarray}
To cross check this conclusion that the ratios are insensitive to new physics contribution,
we presented the $R_\mu$ and $R_\tau$ as functions of $q^2$ in
Fig.~\ref{Fig:6}, where the solid (green), and dotted (blue), dashed
(red) and dot-dashed (orange) lines represent results from the
standard model, S1, S2 and ELV parameters, respectively. We show
that $R_\l (\l=\mu,\tau)$ are  almost unchanged, so that they are not suitable for
searching for $Z^\prime$ effects. These results confirm the conclusion in
Ref.\cite{Hatanaka:2008gu}.
\end{itemize}
\section{summary}
A new family non-universal $Z^\prime$ boson could be naturally
derived in many extensions of SM. One of the possible way to get
such non-universal $Z^\prime$ boson is to include an addition
$U^{\prime}(1)$ gauge symmetry, which has been studied by many
groups. With the data, people had fitted two sets of coupling constants,
S1 and S2 namely. In this work, we have considered the contributions
of family non-universal $Z^\prime$ model  at the tree level in
semi-leptonic $B$ decays involving axial-vector meson $\kone$ in the
final states. The strange axial-vector mesons, $K_1(1270)$ and
$K_1(1400)$, are the mixtures of the $K_{1A}$ and $K_{1B}$, which
are the $1^3P_1$ and $1^1P_1$ states, respectively. We show that the
mixing angle could be constrained by measuring some observables of
$B \to \koneh \lpm$, such as the normalized differential
forward-backward asymmetry and longitudinal lepton polarization
asymmetry. With $\theta=-34 \degree$, the branching ratio of $B \to
\konel \mupm$ is enhanced about by $50\%(30\%)$ with respect to the
corresponding SM values by $Z^\prime$ in S1 (S2). We also found FBA
and lepton polarization asymmetries show quite significant
discrepancies with respect to the SM values. The zero crossing
position in the FBA spectrum at low dilepton mass will move to the
positive direction with $Z^\prime$ boson contribution. We also note
that $B \to \koneh \mupm$ is not suitable to probe new physics,
which will be buried by the uncertainty from the mixing angle. While
for the tauon modes, the new physics contributions are not
remarkable due to small phase spaces except in the extreme limit.
These results could be tested in the running LHC-b experiment and
designed Super-B factory.
\section*{Acknowledgement}
The work of Y. Li is supported in part by the NSFC ((Nos.10805037
and 11175151)) and the Natural Science Foundation of Shandong
Province (ZR2010AM036). K. C. Y. is supported in part by the
National Center for Theoretical Sciences and the National Science
Council of R.O.C. under Grant No. NSC99-2112-M-003-005-MY3.
\begin{appendix}
\section*{Appendix}
\begin{eqnarray}
\A^\kone(\hats) &=& \frac{2}{1+\hatm_{K_1}} C_9^{\eff} (\hats)
A^\kone(\hats) + \frac{4\hatm_b}{\hats} C_7^\eff T^\kone_1(\hats),
\label{Eq:A}
\\
\B^\kone(\hats) &=& (1+\hatm_{K_1})\left[
 C_9^\eff (\hats) V_1^\kone(\hats)
 + \frac{2\hatm_b}{\hats} (1-\hatm_{K_1})C_7^\eff T^\kone_2(\hats)
\right],
\\
\C^\kone(\hats) &=& \frac{1}{1-\hatm_\kone^2} \left[
 (1-\hatm_{K_1}) C_9^\eff(\hats) V_2^\kone (\hats) + 2\hatm_b C_7^\eff
 \left(
  T_3^\kone(\hats) + \frac{1-\hatm_\kone^2}{\hats} T_2^\kone(\hats)
 \right)
\right],
\nonumber\\
\\
\D^\kone(\hats) &=& \frac{1}{\hats} \biggl[
 C_9^\eff(\hats) \left\{(1+\hatm_\kone) V_1^\kone(\hats)
  - (1-\hatm_\kone) V_2^\kone(\hats)
  - 2\hatm_\kone V_0^\kone(\hats) \right\}
  - 2\hatm_b C_7^\eff T_3^\kone(\hats)
\biggr],
\\
\E^\kone(\hats) &=& \frac{2}{1+\hatm_\kone} C_{10} A^\kone(\hats),
\\
\F^\kone(\hats) &=& (1 + \hatm_\kone) C_{10} V_1^\kone(\hats),
\\
\G^\kone(\hats) &=& \frac{1}{1 + \hatm_\kone}C_{10}
V_2^\kone(\hats),
\\
\H^\kone(\hats) &=& \frac{1}{\hats} C_{10} \left[
 (1+\hatm_\kone) V_1^\kone(\hats)
 - (1-\hatm_{K_1}) V_2^\kone(\hats) - 2\hatm_\kone V_0^\kone(\hats)
\right]. \label{Eq:H}
\end{eqnarray}
\end{appendix}


\begin{thebibliography}{99}
\bibitem{Hatanaka:2008gu}
  H.~Hatanaka and K.~C.~Yang,
  Phys.\ Rev.\  D {\bf 78}, 074007 (2008)
  [arXiv:0808.3731 [hep-ph]].


\bibitem{Li:2009rc}
  R.~H.~Li, C.~D.~Lu and W.~Wang,
  Phys.\ Rev.\  D {\bf 79}, 094024 (2009)
  [arXiv:0902.3291 [hep-ph]].

\bibitem{Paracha:2007yx}
  M.~A.~Paracha, I.~Ahmed and M.~J.~Aslam,
  Eur.\ Phys.\ J.\  C {\bf 52}, 967 (2007)
  [arXiv:0707.0733 [hep-ph]].

\bibitem{Bashiry:2009wq}
  V.~Bashiry,
  JHEP {\bf 0906}, 062 (2009)
  [arXiv:0902.2578 [hep-ph]].

\bibitem{Ahmed:2008ti}
  I.~Ahmed, M.~A.~Paracha and M.~J.~Aslam,
  Eur.\ Phys.\ J.\  C {\bf 54}, 591 (2008)
  [arXiv:0802.0740 [hep-ph]];\\
  A.~Saddique, M.~J.~Aslam and C.~D.~Lu,
  Eur.\ Phys.\ J.\  C {\bf 56}, 267 (2008)
  [arXiv:0803.0192 [hep-ph]];\\
  I.~Ahmed, M.~A.~Paracha and M.~J.~Aslam,
  Eur.\ Phys.\ J.\  C {\bf 71}, 1521 (2011)
  [arXiv:1002.3860 [hep-ph]].


\bibitem{Bashiry:2009wh}
  V.~Bashiry and K.~Azizi,
  JHEP {\bf 1001}, 033 (2010)
  [arXiv:0903.1505 [hep-ph]].

\bibitem{Ahmed:2011vr}
  A.~Ahmed, I.~Ahmed, M.~A.~Paracha and A.~Rehman,
  arXiv:1105.3887 [hep-ph].

\bibitem{Suzuki:1993yc}
  M.~Suzuki,
  Phys.\ Rev.\  D {\bf 47}, 1252 (1993).

\bibitem{Burakovsky:1997ci}
  L.~Burakovsky and J.~T.~Goldman,
  Phys.\ Rev.\  D {\bf 57}, 2879 (1998)
  [arXiv:hep-ph/9703271].

\bibitem{Cheng:2003bn}
  H.~Y.~Cheng,
  Phys.\ Rev.\  D {\bf 67}, 094007 (2003)
  [arXiv:hep-ph/0301198].

\bibitem{Hatanaka:2008xj}
  H.~Hatanaka and K.~C.~Yang,
  Phys.\ Rev.\  D {\bf 77}, 094023 (2008)
  [arXiv:0804.3198 [hep-ph]].

\bibitem{Dag:2010jr}
  H.~Dag, A.~Ozpineci and M.~T.~Zeyrek,
  J.\ Phys.\ G {\bf 38}, 015002 (2011)
  [arXiv:1001.0939 [hep-ph]];\\
 M.~Bayar and K.~Azizi,
  Eur.\ Phys.\ J.\  C {\bf 61}, 401 (2009)
  [arXiv:0811.2692 [hep-ph]].

\bibitem{Yang:2008xw}
  K.~C.~Yang,
  Phys.\ Rev.\ D {\bf 78}, 034018 (2008)
  [arXiv:0807.1171 [hep-ph]].

\bibitem{Li:2009tx}
  R.~H.~Li, C.~D.~Lu and W.~Wang,
  Phys.\ Rev.\  D {\bf 79}, 034014 (2009)
  [arXiv:0901.0307 [hep-ph]].

\bibitem{Cheng:2009ms}
  H.~Y.~Cheng and C.~K.~Chua,
  Phys.\ Rev.\  D {\bf 81}, 114006 (2010)
   [arXiv:0909.4627 [hep-ph]];\\
  R.~C.~Verma,
  arXiv:1103.2973 [hep-ph].


\bibitem{Buchalla:1995dp}
  G.~Buchalla, G.~Burdman, C.~T.~Hill and D.~Kominis,
  Phys.\ Rev.\  D {\bf 53}, 5185 (1996)
  [arXiv:hep-ph/9510376].

\bibitem{Nardi:1992nq}
  E.~Nardi,
  Phys.\ Rev.\  D {\bf 48}, 1240 (1993)
  [arXiv:hep-ph/9209223].

\bibitem{Langacker:2000ju}
  P.~Langacker and M.~Plumacher,
  Phys.\ Rev.\  D {\bf 62}, 013006 (2000)
  [arXiv:hep-ph/0001204].

\bibitem{Barger:2009hn}
  V.~Barger, {\sl et. al},
  Phys.\ Lett.\  B {\bf 580}, 186 (2004)
  [arXiv:hep-ph/0310073];\\
  V.~Barger, {\sl et. al},
  Phys.\ Lett.\  B {\bf 598}, 218 (2004)
  [arXiv:hep-ph/0406126];\\
  V.~Barger, {\sl et. al},
  arXiv:0906.3745 [hep-ph];\\
 V.~Barger, {\sl et. al},
  Phys.\ Rev.\  D {\bf 80}, 055008 (2009)
  [arXiv:0902.4507 [hep-ph]].
\bibitem{Cheung:2006tm}
  K.~Cheung, {\sl et. al},
  Phys.\ Lett.\  B {\bf 652}, 285 (2007)
  [arXiv:hep-ph/0604223];\\
  C.~W.~Chiang, {\sl et. al},
  JHEP {\bf 0608}, 075 (2006)
  [arXiv:hep-ph/0606122];\\
  C.~H.~Chen and H.~Hatanaka,
  Phys.\ Rev.\  D {\bf 73}, 075003 (2006)
  [arXiv:hep-ph/0602140];\\
  C.~H.~Chen,
  arXiv:0911.3479 [hep-ph];\\
  C.~W.~Chiang, R.~H.~Li and C.~D.~Lu,
  arXiv:0911.2399 [hep-ph];\\
  R.~Mohanta and A.~K.~Giri,
  Phys.\ Rev.\  D {\bf 79}, 057902 (2009)
  [arXiv:0812.1842 [hep-ph]];\\
  J.~Hua, C.~S.~Kim and Y.~Li,
  Phys.\ Lett.\  B {\bf 690}, 508 (2010)
  [arXiv:1002.2532 [hep-ph]];\\
  J.~Hua, C.~S.~Kim and Y.~Li,
  Eur.\ Phys.\ J.\  C {\bf 69}, 139 (2010)
  [arXiv:1002.2531 [hep-ph]].

\bibitem{Chang:2009wt}
  Q.~Chang, X.~Q.~Li and Y.~D.~Yang,
  JHEP {\bf 0905}, 056 (2009)
  [arXiv:0903.0275 [hep-ph]];\\
    Q.~Chang, X.~Q.~Li and Y.~D.~Yang,
  JHEP {\bf 1002}, 082 (2010)
  [arXiv:0907.4408 [hep-ph]];\\
  Q.~Chang, X.~Q.~Li and Y.~D.~Yang,
  JHEP {\bf 1004}, 052 (2010)
  [arXiv:1002.2758 [hep-ph]];\\
  Q.~Chang and Y.~H.~Gao,
  Nucl.\ Phys.\  B {\bf 845}, 179 (2011)
  [arXiv:1101.1272 [hep-ph]].

\bibitem{Langacker:2008yv}
  P.~Langacker,
  arXiv:0801.1345 [hep-ph].

\bibitem{new8} J.Dickens, V.Gibon, C.Lazzeroni and M.Patel,
CERN-LHCB-2007-038;\\
 J.Dickens, V.Gibon, C.Lazzeroni and M.Patel,
CERN-LHCB-2007-039.

\bibitem{new9} 
  B.~Aubert {\it et al.}  [BABAR Collaboration],
  Phys.\ Rev.\ Lett.\  {\bf 91}, 221802 (2003)
  [arXiv:hep-ex/0308042];\\
  B.~Aubert {\it et al.}  [BABAR Collaboration],
  Phys.\ Rev.\  D {\bf 73}, 092001 (2006)
  [arXiv:hep-ex/0604007];\\
 B.~Aubert {\it et al.}  [BABAR Collaboration],
  Phys.\ Rev.\  D {\bf 79}, 031102 (2009)
  [arXiv:0804.4412 [hep-ex]];\\
  I.~Adachi {\it et al.}  [Belle Collaboration],
  arXiv:0810.0335 [hep-ex].

\bibitem{Bettler:2009gt}
  M.~O.~Bettler  [LHCb Collaboration],
  arXiv:0910.0942 [hep-ex].

\bibitem{PDG}
K. Nakamura et al. (Particle Data Group), J. Phys. G 37, 075021.
(2010)

\bibitem{Buchalla:1995vs}
  G.~Buchalla, A.~J.~Buras and M.~E.~Lautenbacher,
  Rev.\ Mod.\ Phys.\  {\bf 68}, 1125 (1996)
  [arXiv:hep-ph/9512380].

\bibitem{b to s in theory}
  C.S. Kim, T. Morozumi, A.I. Sanda, Phys. Lett. B {\bf 218} (1989)
  343;\\
  X.~G.~He, T.~D.~Nguyen andR.~R.~Volkas,
  Phys.\ Rev.\  D {\bf 38} (1988) 814;\\
  B. Grinstein, M.J. Savage, M.B. Wise, Nucl. Phys. B {\bf 319} (1989)
  271;\\
  N.~G.~Deshpande, J.~Trampetic and K.~Panose,
  Phys.\ Rev.\  D {\bf 39} (1989) 1461;\\
  P.~J.~O'Donnell and H.~K.~K.~Tung,
  Phys.\ Rev.\  D {\bf 43} (1991) 2067;\\
  N.~Paver and Riazuddin,
  Phys.\ Rev.\  D {\bf 45} (1992) 978;\\
  A. Ali,  T. Mannel and T. Morozumi, Phys. Lett.  B{\bf 273} (1991)
  505;\\
  D.~Melikhov, N.~Nikitin and S.~Simula,
  Phys.\ Lett.\  B {\bf 430} (1998) 332  [arXiv:hep-ph/9803343];\\
  J.~M.~Soares, Nucl.\ Phys.\  B {\bf 367} (1991)575;\\
  G.~M.~Asatrian and A.~Ioannisian,
  Phys.\ Rev.\  D {\bf 54} (1996) 5642
  [arXiv:hep-ph/9603318].

\bibitem{Buras:1993xp}
  A.~J.~Buras, M.~Misiak, M.~Munz and S.~Pokorski, Nucl.\ Phys.\  B {\bf 424} (1994) 374 [hep-ph/9311345].

\bibitem{resEff}
 M.~Beneke, G.~Buchalla, M.~Neubert, C.~T.~Sachrajda, Eur.\ Phys.\ J.\ C {\bf 61} (2009) 439 arXiv:0902.4446
 [hep-ph];\\
 M.~Bartsch, M.~Beylich, G.~Buchalla, D.~N.~Gao, JHEP {\bf 0911} (2009) 011 [arXiv:0909.1512
 [hep-ph]];\\
  A.~Khodjamirian, T.~Mannel, A.~A.~Pivovarov and Y.~M.~Wang,
  JHEP {\bf 1009}, 089 (2010)
  [arXiv:1006.4945 [hep-ph]].


\bibitem{Chetyrkin:1996vx}
  K.~G.~Chetyrkin, M.~Misiak and M.~Munz, Phys.\ Lett.\  B {\bf 400} (1997) 206 [Erratum-ibid.\  B {\bf 425} (1998) 414] [hep-ph/9612313].

\bibitem{Altmannshofer:2008dz}
  W.~Altmannshofer, P.~Ball, A.~Bharucha, A.~J.~Buras, D.~M.~Straub and M.~Wick, JHEP {\bf 0901} (2009) 019 [arXiv:0811.1214 [hep-ph]].

\bibitem{UTfit}
 M.~Bona {\it et al.}, arXiv:0906.0953 [hep-ph];\\
 M.~Bona {\it et al.}  [UTfit Collaboration],
  PMC Phys.\  A {\bf 3}, 6 (2009)
  [arXiv:0803.0659 [hep-ph]].


\end{thebibliography}
\end{document}